\documentclass{Interspeech2024}
\usepackage{bbding} 
\usepackage{makecell}

\interspeechcameraready

\title{Leveraging Language Model Capabilities for Sound Event Detection}

\name[affiliation={1,2}]{Hualei}{Wang}
\name[affiliation={1,2}]{Jianguo}{Mao}
\name[affiliation={1,2}]{Zhifang}{Guo}
\name[affiliation={3}]{Jiarui}{Wan}
\name[affiliation={1,2}]{Hong}{Liu}
\name[affiliation={1,2,*}]{Xiangdong}{Wang} 

\address{
  $^1$Beijing Key Laboratory of Mobile Computing and Pervasive Device, Institute of Computing Technology, Chinese Academy of Sciences, Beijing, China \\
  $^2$University of Chinese Academy of Sciences, Beijing, China \\
  $^3$Beijing Jiaotong University, Beijing, China}
\email{\{wanghualei23s, maojianguo20s, guozhifang21s\}@ict.ac.cn wanjiarui68@gmail.com \\\{hliu, xdwang\}@ict.ac.cn}

\keywords{sound event detection, language model, data augmentation}

\begin{document}

\maketitle
\renewcommand{\thefootnote}{\fnsymbol{footnote}}
\footnotetext{*Corresponding author.}

\begin{abstract}
Large language models reveal deep comprehension and fluent generation in the field of multi-modality. Although significant advancements have been achieved in audio multi-modality, existing methods are rarely leverage language model for sound event detection (SED). In this work, we propose an end-to-end framework for understanding audio features while simultaneously generating sound event and their temporal location. Specifically, we employ pretrained acoustic models to capture discriminative features across different categories and language models for autoregressive text generation. Conventional methods generally struggle to obtain features in pure audio domain for classification. In contrast, our framework utilizes the language model to flexibly understand abundant semantic context aligned with the acoustic representation. The experimental results showcase the effectiveness of proposed method in enhancing timestamps precision and event classification.
\end{abstract}

\section{Introduction}
Sounds contain considerable information about the environment. Sound event detection (SED) aims to detect and classify various sound events within an audio stream. Unlike other acoustic tasks such as tagging and caption, SED requires improved discrimination to accurately identify the temporal boundaries of occurring events. SED has broad applications in human-machine interaction~\cite{yin2021wearable} and automatic sound recognition~\cite{sigtia2016automatic}. It can also provide valuable benefits in the fileds of smart homes~\cite{debes2016monitoring} and smart cites~\cite{bello2018sound}. 

\begin{figure}[!t]
    \centering
    \includegraphics[width=\linewidth]{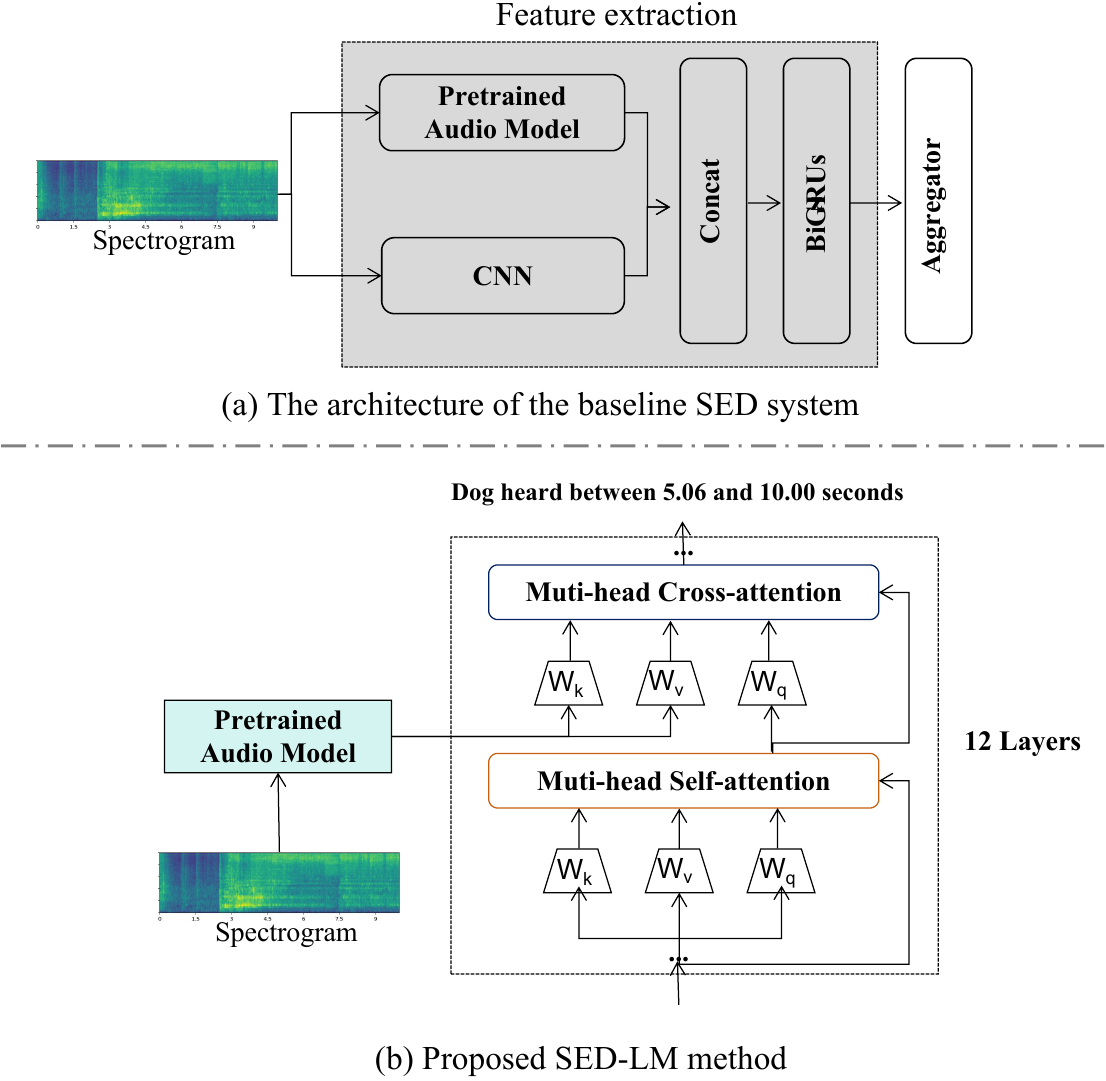}
    \caption{We represent the DCASE2023 baseline SED systems and language model for automatic SED. (a) shows baseline SED system. (b) is the architecture of SED-LM. The module in the dashed block is a transformer layer with cross-attention.}
    \label{model}
\end{figure}

The general paradigm of SED primarily concentrates on classification of frame-wise features extracted from the audio. Designing an effective feature extractor to identify specific events at the frame level has revealed its importance in recent years. The Convolutional Recurrent Neural Network (CRNN)~\cite{cakir2017convolutional} is commonly utilized as fundamental framework to learn high level features from the audio spectrogram. Where data are only labeled with event categories rather than both categories and boundaries, an aggregator is employed to pool the frame-level features, generated by the feature extractor, into a clip-level feature for the joint training of weakly labeled data~\cite{lin2020specialized}. Inspired by the success of the self-supervised learning (SSL) and pretrained models in the fields of text and image processing, the trend of adopting pretrained acoustic models is gradually becoming mainstream. Figure~\ref{model} (a) presents the architecture of the baseline system for the DCASE 2023 Task 4 challenge, incorporating a pre-trained audio model (BEATs) as an additional feature extractor to yield highly discriminative embeddings for improved performance. 

In recent years, pretrained language models such as BERT~\cite{devlin2018bert}, BART~\cite{lewis2019bart} and T5~\cite{raffel2020exploring} have exhibited remarkable capabilities in natural language understanding and reasoning~\cite{radford2019language,dai2019transformer,raffel2020exploring,zhang2022opt}. Researchers have also deeply explored the language models in the mutimodal domain~\cite{tsimpoukelli2021multimodal}, recognizing that semantic knowledge acquired by language models can benefit other modalities. The integration of acoustic and language modalities aims to build a common learning, where each modality mutually enhance the capabilities of the other. WavCaps~\cite{mei2023wavcaps} employs ChatGPT to generate a large-scale and diverse captioning dataset, releasing potential abilities of audio-language multimodal learning. Under the guidance of natural language, Contrastive Language Audio Pre-training (CLAP)~\cite{wu2022wav2clip,elizalde2023clap} integrates natural language and audio into a unified multimodal space using dual encoders and contrastive learning, enabling zero-shot capabilities without the need for specific acoustic class labels. Integrating conventional audio models with large language models has proven effective not only for sound perception and reasoning but also for application in audio classification and retrieval tasks~\cite{gong2023listen,deshmukh2024pengi}. APT-LLM~\cite{liang2023acoustic} aims to train a model to recognise specific sound events and count their frequency. However, based on audio-language model have been rarely explored for SED due to challenges in representing, aligning, and generating frame-wise event boundaries in the text modality. To solve these problems, we select different architectures to thoroughly leverage the multimodal capabilities for end-to-end prediction of SED. 

In this paper, we propose a novel method referred to as SED-LM that leverages the capabilities of language models to uniquely generate text describing sound event and their temporal aspects. The audio embeddings and the language embeddings are integrated through cross-attention mechanism. Autoregressive generation is utilized during the inference stage. In this way, SED-LM gains the capability to distinguish sound events more easily from textual representation.

Our contributions can be specified as follows:
\begin{itemize}
\item We propose an end-to-end method to generate SED content in multi-modal, which is flexiblely adapted to different audio feature extractor component and the pretrained language generator.
\item Our method combines the strengths of pretrained language models with audio models, achieving multi-modal integration through cross-attention mechanism. Language model generates text corresponding to relevant audio segments with precise occurrence of sound event.
\item Extensive Experiments are conducted to evaluate the interaction between the audio and language models, as well as the sampling strategy employed. The results of the BEATs with BERT model achieves 0.607 EB-F1 score with supervised learning, which showcase the effectiveness of the synergy between audio and language models in SED.
\end{itemize}

\section{Methodology}
To guarantee that each audio is paired with text, we select annotated training datasets. In this section, we briefly introduce overview of our proposed methodology. Then, we present the audio extraction and decoder in detail.


\subsection{Overview of proposed SED-LM model}
Single-modal information is inherently limited, multimodal approaches yield more comprehensive insights for SED tasks. Rather than using the homogeneous audio feature, our proposed method generates sound event and their temporal location with language model and introduces a cross-modal perspective for SED feature learning. The method that leverages semantic information from language model is capable of obtaining fine-grained audio features. But integrating cross-modal audio and text for SED presents some difficulties: 1)audio embeddings need to be aligned with variable-length text embeddings. 2)selecting appropriate text sequences is hard for SED-LM. 

Therefore, the architecture of our proposed model is illustrated in Figure~\ref{model} (b), which consists of a pretrained audio encoder ${Enc^A}$ and a text decoder ${Dec^T}$. The audio encoder ${Enc^A}$ processes acoustic feature. ${Dec^T}$ is composed of 12 transformer layers, each equipped with a self-attention module followed by a cross-attention module with a residual connection. The integration of text and audio modalities is facilitated through cross-attention, where acoustic embeddings serve as key/value pairs and text hidden states serve as the query. 
\begin{table}[h]
  \caption{templates and enhanced text. "/" refers to \textless SEP\textgreater.}
  \label{tab:table0}
  \centering
  \resizebox{\linewidth}{!}{
    \begin{tabular}{|c|c|}
        \hline
         alarm 1.5 3.5 / speech 7.1 9.2 & speech 7.1 9.2 / alarm 1.5 3.5 \\ \hline
         \makecell {cat heard 2.2 5.5 seconds / \\ dog heard 3.3 4.7 seconds} & \makecell {dog heard 3.3 4.7 seconds / \\ cat heard 2.2 5.5 seconds} \\ \hline
         alarm heard between 1.5 and 3.5 seconds & - \\ \hline
    \end{tabular}}
\end{table}
To train SED-LM model, we exploit text templates rather than original (${s, e, C}$) labels, where ${s, e}$ separately represents the start and end timestamps and $C$ denotes the sound event category. Given different template extremely influence the generative effect, we select texts of varying types, including (C s e), (C heard s e seconds) and (C heard between s and e seconds). Then we transform text to discrete labels and employed a pattern matching recognition method to extract temporal information and sound class from the prediction sequences. The templates are shown in Table \ref{tab:table0}.

\subsection{Generative SED}
\subsubsection{Audio extraction: from spectrograms to features}
The audio encoder takes advantage of the pre-trained large model for adaptation to the SED. We compare two audio extractors.
The first HTS-AT~\cite{chen2022hts} employs transformer-based architectures as its backbone. It contains four groups of transformer-encoder blocks with window attention mechanism and patch-merge layers. The input spectrograms are divided into patch tokens, covering 64 frequency bands by 64 frames in each window, then the shape of the patch tokens is reduced by 8 times following network groups. The encoder transforms the feature into an embedding $E^A\in R^{T\ \times D}$. Although HTS-AT has shown superior feature extraction, we also employ BEATs to verify the robustness of architecture and compare performance aligned with language model. The pretrained BEATs, featuring 12-layer transformer encoder, is optimized through iterative pre-training of the acoustic tokenizer and audio SSL model. After training, the audio feature token contains information related to corresponding time frames and frequency bands. Both BEATs and HTS-AT require parameters fine-tuning for effective integration with the language model. 

\subsubsection{Language models: from audio tokens to generated text}
The decoder module takes into account the differences in architecture between masked and generative type. BERT model is typically pretrained on extensive masked text corpora, it is naturally designed for masked language tasks. The bidirectional self-attention mechanism is not suitable for generation tasks. Therefore, we adapt the structure to a unidirectional self-attention mechanism that only considers past context. Compared to BERT, BART incorporates both encoder and decoder blocks within a sequence-to-sequence architecture. We specifically employ only the decoder part of BART. T5 model adopts a unified text-to-text approach regardless of the task and is utilized in a manner similar to BART. 

The decoder is trained with template texts for SED understanding and generating based on the previously generated text. During model training, we apply language modeling loss for text decoder, which utilizes a forward autoregressive formulation to learn conditional probability of the generated text. During inference, the model produces texts autoregressively using only the audio input.
\begin{align}
\label{eq:audio encoder}
E^A = \text{$Enc$}^A(X)
\end{align}

\begin{align}
\label{eq:decoder}
y = \text{$Dec$}^T(E^A)
\end{align}
\begin{align}
\label{eq:loss}
\mathcal{L}_{\mathrm{ce}}(\theta)=-\frac{1}{T}\sum_{t=1}^{T}\log P_\theta(y_t|y_{1:t-1},E^A)
\end{align}
where $X$ represents the input audio features, $y_t$ denotes the $t$-th ground truth token in a sentence of length $T$, and $\theta$ represents the model parameters.

\subsubsection{Cross attention: integrate audio and text}
Additionally, cross-attention is added to connect the audio encoder and the language decoder in each layer, enhancing the model's ability to integrate frame and text token. The current state of the decoder is $E^q\in R^{N\ \times D}$. The cross-attention mechanism is formulated as:

\begin{align}
\label{eq:cross-attention}
\text{Cross-attention}(E^q,E^A)=\left(\frac{(W^qE^q)(W^kE^A)^{\rm T}}{\sqrt{d}}\right)W^vE^A
\end{align}

\subsubsection{Text augmentation}
Text augmentation is applied for the generative text. The text describing sound events reflects the temporal sequence of audio. Appropriate augmentation of the dependent texts assists the model in learning and modeling multiple sound events, accurately capturing temporal relationships. Text augmentation effectively address the issue of overfitting during training. 

\section{Experiments Setup}

\subsection{Datasets}
Audioset~\cite{gemmeke2017audio}, the largest weakly-annotated sound event dataset, contains over 2M 10-sec sound segments. AudioSet provides a small-scale strongly-annotated subset which contains additional on-sets and off-sets of present events. All experiments are conducted using the dataset of the DCASE 2023 task 4~\cite{turpault2019sound}, which contains the supervised data for weakly labeled data (1578 clips) and synthetic labeled set (10000 clips) created by soundscape synthesis tool in the domestic environment. Extra real strong labeled set (3470 clips) from the AudioSet are used as well. The real validation data (1168 clips) is prepared for evaluation. 

\subsection{Experimental Settings}
In pre-processing, the sampling rate of all input audio clips is converted to 16k. The audio clips are random cropped or padding to 10 seconds. The HT-SAT model use 1024 window size that shifts every 10 ms and 64 mel-bins to computed spectrograms. As for the BEATs, we use 128-dimensional Mel-filter bank features with 400 window size. For language models, the BERT base model contains 110M parameters, BART base is equipped with 140M parameters, and T5 also has substantial 220M parameters. The max length of text sequence is confined to 256 tokens.
The batch size for audio-text paired is 4. The learning rate is warm-up to $1e^{-4}$ in the first 5000 iterations, and decayed to $1e^{-5}$ with cosine schedule. We use the AdamW optimizer with a weight decay of 0.05. The pretrained model is further fine-tuned for 15 epochs. 

\subsection{Evaluation metrics}
Collarbased F1 score (EB-F1), intersection-based F1 score (IB-F1) are used to evaluate the model performance~\cite{mesaros2016metrics}. EB-f1 score is designed to validate the accuracy of sound onsets and offsets. Therefore, former metric focuses more on the outcomes of continuous SED, while the latter metric emphasizes the precision of audio tagging. The metric combines EB-F1 and IB-F1 for comprehensive assessment. Moreover, the model is evaluated using the validation of the DCASE2023 Challenge Task4. 

\begin{figure}
  \begin{minipage}[t]{0.49\linewidth}
    \centering
    \includegraphics[scale=0.517]{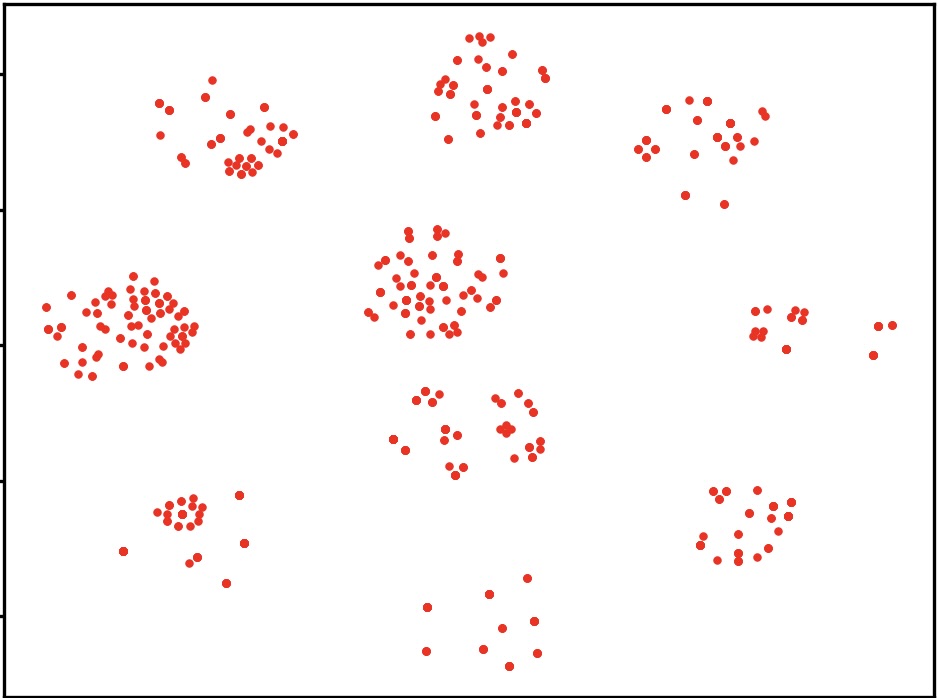}
  \end{minipage}%
  \hspace{2.835pt}
  \begin{minipage}[t]{0.48\linewidth}
    \centering
    \includegraphics[scale=0.482]{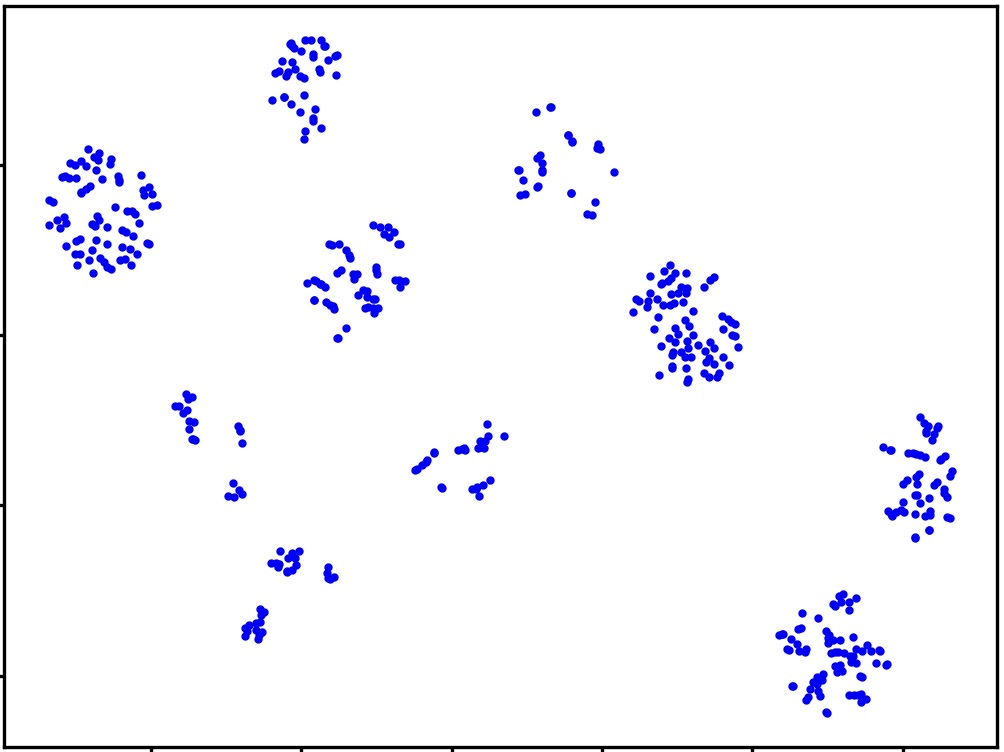}
  \end{minipage}
  \caption{The trained HT-SAT and BEATs feature}
  \label{fig:feature}
\end{figure}

\section{Experiments Result}
We firstly evaluate our SED-LM over temporal accuracy and tagging classification compared to single modal models. Then, we conduct experiment to analyse the generative prefermance with different module. In the end, we explore enhanced strategies in text augmentation and beam size.

\subsection{Comparison between proposed and other method for SED}

We compare the performance between the baseline and a few single modal methods. The conventional methods are trained on the whole DCASE2023 datasets, while our proposed method merely depend on labeled and weakly labeled data.
The results are shown in Table \ref{tab:table1}. For DCASE2023 baseline system, the BEATs parameters are frozen. The first two single modal methods are designed to adjust the kernel and convolution. The AST-SED and ATST-SED obtain superior performance utilizing pretrained model.
The novel SED-LM can outperform the baseline and most single modal models, which proves generative method qualifies in SED. As the training dataset excludes unlabeled data (14412 clips), it is not as good as the state-of-the-art method.
The semantic information conveyed in the generated content correlates to SED. This is consistent with the characteristics of multi-modal.

\subsection{Integrated feature analysing}
To explore model’s ability to understand acoustic spectral information at a fine-grained level. We visualized feature of trained acoustic layer with the TSNE~\cite{van2008visualizing} technique. For both models, we randomly select one frame-level feature from every real audio clip in the validation dataset and plot them. The visualization results are shown in Figure \ref{fig:feature}. Comparing the features of HT-SAT and BEATs, we can see that most of the acoustic features modeled are clearly discriminative. Moreover, the BEATs features are more densely clustered comparing with the HTSAT features, illustrating an obvious separable property. 

\begin{figure}[!t]
\centering
    \includegraphics[width=\linewidth]{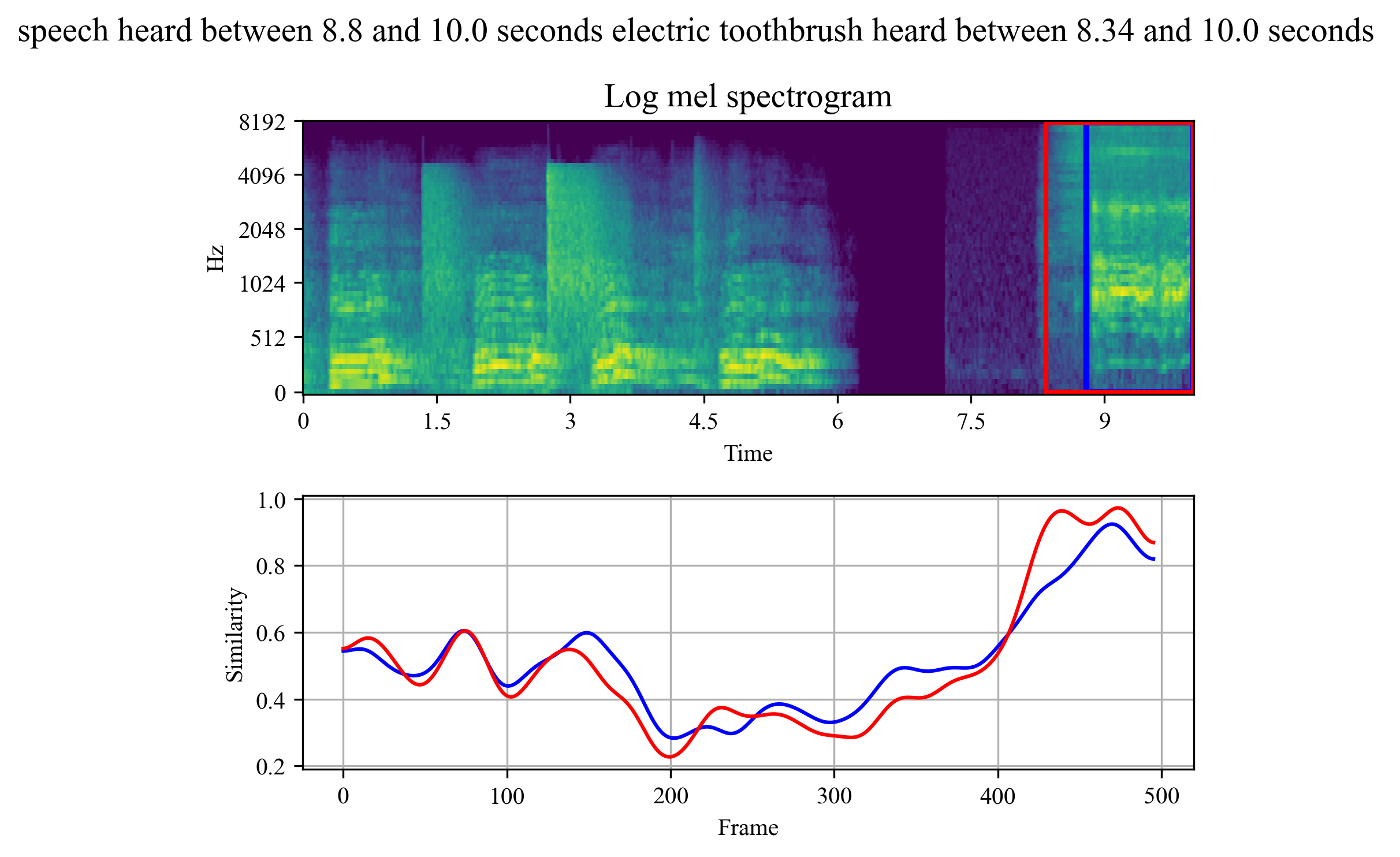}
\caption{Y0s6Vt4i3iiQ\_40.000\_50.000.wav's similarity. The electric toothbrush event's similarity between frame and corresponding text is represented by red color. The speech is represented by blue color.}
\label{fig:simi}
\end{figure}

Specifically, Figure \ref{fig:simi} gives an example in which text contexts express sound detection. The similarity values of the audio features and text feature are higher than those of the noise clip during speech and electric toothbrush events. This modeling way with cross attention mechanism align the resemblance of fine-grained audio and the text feature. The cross-attention mechanism integrate representations that relate to both the continuity of the audio and occurrence of events.

\subsection{Ablation study of the proposed SED-LM paradigm}
\begin{table}[h]
  \caption{Comparison of model performance on DCASE 2023 task4. Our proposed method merely depend on labeled datasets.}
  \label{tab:table1}
  \centering
    \begin{tabular}{ccccc}
        \toprule
         \textbf{Model} & \textbf{EB-F1(\%)} $\uparrow$ & \textbf{IB-F1(\%)} $\uparrow$\\
        \midrule
        \midrule
        CRNN-SED & 42.35 & 64.22 \\
        \midrule
        FDY-CRNN~\cite{nam2022frequency} & 51.56  & 74.50 \\
        SK-CRNN~\cite{zheng2021improved} & 52.77  &  74.88\\
        AST-SED~\cite{li2023ast} & 59.60  &  79.64\\
        ATST-SED~\cite{shao2023fine} & 65.83  &  $-$\\
        SED-LM (proposed) &  60.72 &  80.60 \\
        \bottomrule 
    \end{tabular}
\end{table}

\subsubsection{Template texts}
Different template texts and their performance are detailed in Table \ref{tab:table2},  showcasing that the final one yields better results in the generative system. Complete and logically coherent texts are more conducive for language models learning.

\begin{table}[h]
  \caption{Ablation study of different text template.}
  \label{tab:table2}
  \centering
  \resizebox{\linewidth}{!}{
    \begin{tabular}{ccc}
        \toprule
         \textbf{Text} & \textbf{EB-F1(\%)} & \textbf{IB-F1(\%)}\\
        \midrule
        \midrule
         \textless C s e\textgreater & 47.5 & 67.6\\ 
         \textless C heard s e seconds\textgreater & 55.6 & 79.0\\
         \textless C heard between s and e seconds\textgreater & 60.7 & 80.6\\
        \bottomrule 
    \end{tabular}}
\end{table}

\subsubsection{Model selection}
As shown in Table \ref{tab:table2}, we explore the influence of the encoder and decoder performance. It can be seen that the effects of three language model are in increasing order.
Firstly, we investigate the effect of the audio encoder in our method. The BEATs model outperforms the HTSAT nearly 5 \%. This is correspond to feature visualization.
Secondly, The results reveal that EB-F1 and IB-F1 of the BERT decoder are significantly higher than those of the BART and T5. 
The BERT decoder selects candidate words with a tendency towards template-based sequences, which generates more discriminating SED content. 
In contrast, the BART and T5 model selects candidate words based on sentence flow, differing from template sequences, which lead to incorrect sound detection generation. 
Then they may not accurately predict the subsequent word and generate poor outcomes in scenarios of multi-events.

\begin{table}
  \caption{Performance of the proposed method on validation metrics within different modules.}
 \label{tab:table3}
 \centering
    \begin{tabular}{cccc}
        \toprule
        \textbf{\makecell{Acoustic \\ encoder}} & \textbf{\makecell{LM \\ decoder}} 
        &\textbf{ \textbf{EB-F1(\%)} $\uparrow$ } &\textbf{ \textbf{IB-F1(\%) $\uparrow$}}\\
        \midrule
        \midrule
        HTS-AT & BART & 40.2 & 62.6 \\
        HTS-AT & T5 & 50.9 & 74.8 \\
        HTS-AT & BERT & 54.2 & 77.4\\
        BEATs & BART & 44.5 & 67.1\\
        BEATs & T5 & 55.4 & 76.0\\
        BEATs & BERT & \textbf{60.7} & \textbf{80.2} \\

        \bottomrule 
    \end{tabular}
\end{table}

\subsubsection{Enhanced strategies}
Table \ref{tab:table4} illustrates the performance of different enhanced strategies, reflecting that the general choice of decoding strategy can significantly impact the quality and accuracy of the generative sequences. 
Both greedy and beam search strategies aim to select the highest probability candidate word at each time step, thereby choosing the most likely token to generate accurate results. 
We observe that greedy search outperform the beam search with 2 and 3 during the generative process.
This can be attributed to BERT's masked pattern, which enables it to assign maximum probability effectively over short ranges.
Text augmentation also enhance SED-LM performance.
\begin{table}[h]
  \caption{Ablation study of the BEATs-BERT enhanced strategy.}
  \label{tab:table4}
  \centering
    \begin{tabular}{cccc}
        \toprule
         \textbf{Augmentation} &  \textbf{Beam size} & \textbf{EB-F1(\%)} & \textbf{IB-F1(\%)}\\
        \midrule
        \midrule
         \Checkmark & 1 (greedy search)& 60.7 & 80.6\\ 
         \Checkmark & 2 & 58.6 & 79.5\\
         \Checkmark & 3 & 57.7 & 78.0\\
         $-$ & 1 (greedy search) & 57.9 & 79.5\\
         $-$ & 2 & 57.5 & 78.7\\
         $-$ & 3 & 56.7 & 78.5\\
        \bottomrule 
    \end{tabular}
\end{table}
\section{Conclusions}
In this paper, we proposed a novel method to generate SED utilizing language models, which harnesses the capability of language models to provide precise SED text. We design an end-to-end architecture that integrate audio and language space. The encoder audio model produces features serving as key and values. Meanwhile, the decoder generate the sound event and temporal location content. The experimental results demonstrate the feasibility of the generative model, which exhibits remarkable generality and multi-modal capability. In addition, from various ablation study, we have investigated the performance of different modules, which achieve comparable results with the single modal.
\bibliographystyle{IEEEtran}
\bibliography{mybib}

\begin{thebibliography}{10}
\providecommand{\url}[1]{#1}
\csname url@samestyle\endcsname
\providecommand{\newblock}{\relax}
\providecommand{\bibinfo}[2]{#2}
\providecommand{\BIBentrySTDinterwordspacing}{\spaceskip=0pt\relax}
\providecommand{\BIBentryALTinterwordstretchfactor}{4}
\providecommand{\BIBentryALTinterwordspacing}{\spaceskip=\fontdimen2\font plus
\BIBentryALTinterwordstretchfactor\fontdimen3\font minus \fontdimen4\font\relax}
\providecommand{\BIBforeignlanguage}[2]{{%
\expandafter\ifx\csname l@#1\endcsname\relax
\typeout{** WARNING: IEEEtran.bst: No hyphenation pattern has been}%
\typeout{** loaded for the language `#1'. Using the pattern for}%
\typeout{** the default language instead.}%
\else
\language=\csname l@#1\endcsname
\fi
#2}}
\providecommand{\BIBdecl}{\relax}
\BIBdecl

\bibitem{yin2021wearable}
R.~Yin, D.~Wang, S.~Zhao, Z.~Lou, and G.~Shen, ``Wearable sensors-enabled human--machine interaction systems: from design to application,'' \emph{Advanced Functional Materials}, vol.~31, no.~11, p. 2008936, 2021.

\bibitem{sigtia2016automatic}
S.~Sigtia, A.~M. Stark, S.~Krstulovi{\'c}, and M.~D. Plumbley, ``Automatic environmental sound recognition: Performance versus computational cost,'' \emph{IEEE/ACM Transactions on Audio, Speech, and Language Processing}, vol.~24, no.~11, pp. 2096--2107, 2016.

\bibitem{debes2016monitoring}
C.~Debes, A.~Merentitis, S.~Sukhanov, M.~Niessen, N.~Frangiadakis, and A.~Bauer, ``Monitoring activities of daily living in smart homes: Understanding human behavior,'' \emph{IEEE Signal Processing Magazine}, vol.~33, no.~2, pp. 81--94, 2016.

\bibitem{bello2018sound}
J.~P. Bello, C.~Mydlarz, and J.~Salamon, ``Sound analysis in smart cities,'' \emph{Computational analysis of sound scenes and events}, pp. 373--397, 2018.

\bibitem{cakir2017convolutional}
E.~Cak{\i}r, G.~Parascandolo, T.~Heittola, H.~Huttunen, and T.~Virtanen, ``Convolutional recurrent neural networks for polyphonic sound event detection,'' \emph{IEEE/ACM Transactions on Audio, Speech, and Language Processing}, vol.~25, no.~6, pp. 1291--1303, 2017.

\bibitem{lin2020specialized}
L.~Lin, X.~Wang, H.~Liu, and Y.~Qian, ``Specialized decision surface and disentangled feature for weakly-supervised polyphonic sound event detection,'' \emph{IEEE/ACM Transactions on Audio, Speech, and Language Processing}, vol.~28, pp. 1466--1478, 2020.

\bibitem{devlin2018bert}
J.~Devlin, M.-W. Chang, K.~Lee, and K.~Toutanova, ``Bert: Pre-training of deep bidirectional transformers for language understanding,'' \emph{arXiv preprint arXiv:1810.04805}, 2018.

\bibitem{lewis2019bart}
M.~Lewis, Y.~Liu, N.~Goyal, M.~Ghazvininejad, A.~Mohamed, O.~Levy, V.~Stoyanov, and L.~Zettlemoyer, ``Bart: Denoising sequence-to-sequence pre-training for natural language generation, translation, and comprehension,'' \emph{arXiv preprint arXiv:1910.13461}, 2019.

\bibitem{raffel2020exploring}
C.~Raffel, N.~Shazeer, A.~Roberts, K.~Lee, S.~Narang, M.~Matena, Y.~Zhou, W.~Li, and P.~J. Liu, ``Exploring the limits of transfer learning with a unified text-to-text transformer,'' \emph{The Journal of Machine Learning Research}, vol.~21, no.~1, pp. 5485--5551, 2020.

\bibitem{radford2019language}
A.~Radford, J.~Wu, R.~Child, D.~Luan, D.~Amodei, I.~Sutskever \emph{et~al.}, ``Language models are unsupervised multitask learners,'' \emph{OpenAI blog}, vol.~1, no.~8, p.~9, 2019.

\bibitem{dai2019transformer}
Z.~Dai, Z.~Yang, Y.~Yang, J.~Carbonell, Q.~V. Le, and R.~Salakhutdinov, ``Transformer-xl: Attentive language models beyond a fixed-length context,'' \emph{arXiv preprint arXiv:1901.02860}, 2019.

\bibitem{zhang2022opt}
S.~Zhang, S.~Roller, N.~Goyal, M.~Artetxe, M.~Chen, S.~Chen, C.~Dewan, M.~Diab, X.~Li, X.~V. Lin \emph{et~al.}, ``Opt: Open pre-trained transformer language models,'' \emph{arXiv preprint arXiv:2205.01068}, 2022.

\bibitem{tsimpoukelli2021multimodal}
M.~Tsimpoukelli, J.~L. Menick, S.~Cabi, S.~Eslami, O.~Vinyals, and F.~Hill, ``Multimodal few-shot learning with frozen language models,'' \emph{Advances in Neural Information Processing Systems}, vol.~34, pp. 200--212, 2021.

\bibitem{mei2023wavcaps}
X.~Mei, C.~Meng, H.~Liu, Q.~Kong, T.~Ko, C.~Zhao, M.~D. Plumbley, Y.~Zou, and W.~Wang, ``Wavcaps: A chatgpt-assisted weakly-labelled audio captioning dataset for audio-language multimodal research,'' 2023.

\bibitem{wu2022wav2clip}
H.-H. Wu, P.~Seetharaman, K.~Kumar, and J.~P. Bello, ``Wav2clip: Learning robust audio representations from clip,'' in \emph{ICASSP 2022-2022 IEEE International Conference on Acoustics, Speech and Signal Processing (ICASSP)}.\hskip 1em plus 0.5em minus 0.4em\relax IEEE, 2022, pp. 4563--4567.

\bibitem{elizalde2023clap}
B.~Elizalde, S.~Deshmukh, M.~Al~Ismail, and H.~Wang, ``Clap learning audio concepts from natural language supervision,'' in \emph{ICASSP 2023-2023 IEEE International Conference on Acoustics, Speech and Signal Processing (ICASSP)}.\hskip 1em plus 0.5em minus 0.4em\relax IEEE, 2023, pp. 1--5.

\bibitem{gong2023listen}
Y.~Gong, H.~Luo, A.~H. Liu, L.~Karlinsky, and J.~Glass, ``Listen, think, and understand,'' \emph{arXiv preprint arXiv:2305.10790}, 2023.

\bibitem{deshmukh2024pengi}
S.~Deshmukh, B.~Elizalde, R.~Singh, and H.~Wang, ``Pengi: An audio language model for audio tasks,'' \emph{Advances in Neural Information Processing Systems}, vol.~36, 2024.

\bibitem{liang2023acoustic}
J.~Liang, X.~Liu, W.~Wang, M.~D. Plumbley, H.~Phan, and E.~Benetos, ``Acoustic prompt tuning: Empowering large language models with audition capabilities,'' 2023.

\bibitem{chen2022hts}
K.~Chen, X.~Du, B.~Zhu, Z.~Ma, T.~Berg-Kirkpatrick, and S.~Dubnov, ``Hts-at: A hierarchical token-semantic audio transformer for sound classification and detection,'' in \emph{ICASSP 2022-2022 IEEE International Conference on Acoustics, Speech and Signal Processing (ICASSP)}.\hskip 1em plus 0.5em minus 0.4em\relax IEEE, 2022, pp. 646--650.

\bibitem{gemmeke2017audio}
J.~F. Gemmeke, D.~P. Ellis, D.~Freedman, A.~Jansen, W.~Lawrence, R.~C. Moore, M.~Plakal, and M.~Ritter, ``Audio set: An ontology and human-labeled dataset for audio events,'' in \emph{2017 IEEE international conference on acoustics, speech and signal processing (ICASSP)}.\hskip 1em plus 0.5em minus 0.4em\relax IEEE, 2017, pp. 776--780.

\bibitem{turpault2019sound}
N.~Turpault, R.~Serizel, A.~P. Shah, and J.~Salamon, ``Sound event detection in domestic environments with weakly labeled data and soundscape synthesis,'' in \emph{Workshop on Detection and Classification of Acoustic Scenes and Events}, 2019.

\bibitem{mesaros2016metrics}
A.~Mesaros, T.~Heittola, and T.~Virtanen, ``Metrics for polyphonic sound event detection,'' \emph{Applied Sciences}, vol.~6, no.~6, p. 162, 2016.

\bibitem{van2008visualizing}
L.~Van~der Maaten and G.~Hinton, ``Visualizing data using t-sne.'' \emph{Journal of machine learning research}, vol.~9, no.~11, 2008.

\bibitem{nam2022frequency}
H.~Nam, S.-H. Kim, B.-Y. Ko, and Y.-H. Park, ``Frequency dynamic convolution: Frequency-adaptive pattern recognition for sound event detection,'' \emph{arXiv preprint arXiv:2203.15296}, 2022.

\bibitem{zheng2021improved}
X.~Zheng, Y.~Song, I.~McLoughlin, L.~Liu, and L.-R. Dai, ``An improved mean teacher based method for large scale weakly labeled semi-supervised sound event detection,'' in \emph{ICASSP 2021-2021 IEEE International Conference on Acoustics, Speech and Signal Processing (ICASSP)}.\hskip 1em plus 0.5em minus 0.4em\relax IEEE, 2021, pp. 356--360.

\bibitem{li2023ast}
K.~Li, Y.~Song, L.-R. Dai, I.~McLoughlin, X.~Fang, and L.~Liu, ``Ast-sed: An effective sound event detection method based on audio spectrogram transformer,'' in \emph{ICASSP 2023-2023 IEEE International Conference on Acoustics, Speech and Signal Processing (ICASSP)}.\hskip 1em plus 0.5em minus 0.4em\relax IEEE, 2023, pp. 1--5.

\bibitem{shao2023fine}
N.~Shao, X.~Li, and X.~Li, ``Fine-tune the pretrained atst model for sound event detection,'' \emph{arXiv preprint arXiv:2309.08153}, 2023.

\end{thebibliography}
\end{document}